\documentclass[%
twocolumn,
 amsmath,amssymb,
 aps,authoryear
]{aastex63}

\newcommand{\add}[1]{{#1}}

\shorttitle{Domains in the solar wind}
\shortauthors{Ruffolo et al.}
\begin{document}
\title{Domains of Magnetic Pressure Balance in Parker Solar Probe Observations of the Solar Wind}
\author[0000-0003-3414-9666]{David Ruffolo}
\correspondingauthor{David Ruffolo}
\email{david.ruf@mahidol.ac.th}
\affiliation{Department of Physics, Faculty of Science, Mahidol University, Bangkok 10400, Thailand}
\author[0000-0002-1794-1427]{Nawin Ngampoopun}
\affiliation{Department of Physics, Faculty of Science, Mahidol University, Bangkok 10400, Thailand}
\author[0000-0001-7710-6557]{Yash R. Bhora}
\affiliation{Wells International School, Bangkok 10260, Thailand}
\author[0000-0001-9597-1448]{Panisara Thepthong}
\affiliation{Department of Physics, Faculty of Science, Kasetsart University, Bangkok 10900, Thailand}
\author[0000-0002-6609-1422]{Peera Pongkitiwanichakul}
\affiliation{Department of Physics, Faculty of Science, Kasetsart University, Bangkok 10900, Thailand}
\author[0000-0001-7224-6024]{William H.~Matthaeus}
\affiliation{Department of Physics and Astronomy, University of Delaware, Newark, DE 19716, USA}
\affiliation{Bartol Research Institute, University of Delaware, Newark, DE 19716, USA}
\author[0000-0002-7174-6948]{Rohit Chhiber}
\affiliation{Department of Physics and Astronomy, University of Delaware, Newark, DE 19716, USA}
\affiliation{Heliophysics Science Division, NASA Goddard Space Flight Center, Greenbelt MD 20771, USA} 

\accepted{by the Astrophysical Journal, 2021 Oct 10}
\begin{abstract}
The {\it Parker Solar Probe (PSP)} spacecraft is performing the first {\it in situ} exploration of the solar wind within 0.2 au of the Sun. 
Initial observations confirmed the Alfv\'enic nature of aligned fluctuations of the magnetic field {\bf B} and velocity {\bf V} in solar wind plasma close to the Sun, in domains of nearly constant magnetic field magnitude $|{\bf B}|$, i.e., approximate magnetic pressure balance.  
Such domains are interrupted by particularly strong fluctuations, including but not limited to radial field (polarity) reversals, known as switchbacks. 
It has been proposed that nonlinear Kelvin-Helmholtz instabilities form near magnetic boundaries in the nascent solar wind leading to extensive shear-driven dynamics, strong turbulent fluctuations including switchbacks, and mixing layers that involve domains of approximate magnetic pressure balance. 
In this work we identify and analyze various aspects of such domains using data from the first five {\it PSP} solar encounters. 
The filling fraction of domains, 
\add{a measure of Alfv\'enicity,} varies from median values of 90\% within 0.2 au to 38\% outside 0.9 au, with strong fluctuations.
We find an inverse association between the mean domain duration and plasma $\beta$.
We examine whether the mean domain duration is also related to the crossing time of spatial structures frozen into the solar wind flow for extreme cases of the aspect ratio.  
Our results are inconsistent with long, thin 
\add{domains} aligned along the radial or Parker spiral direction, 
\add{and compatible with isotropic domains, which is consistent with prior observations of isotropic density fluctuations or ``flocculae'' in the solar wind}.
\end{abstract}
\section{Introduction}

Fluctuating magnetic fields are prevalent in space plasmas.  
In the solar wind, such fluctuations play a key role in the acceleration \citep{Drury83} and transport \citep{Jokipii66} of energetic charged particles, including solar energetic particles, which are an important component of space weather effects of solar storms on human activity \citep{Knipp11}.
Indeed, irregular magnetic fields in the heliosphere were initially inferred from the transport of relativistic solar ions \citep{Sittkus56b,MeyerEA56}, before the solar wind itself was proposed theoretically \citep{Parker58} and confirmed observationally \citep{GringauzEA60,NeugebauerSnyder62}.
Magnetic fluctuations, together with velocity fluctuations, are often found as part of a turbulent cascade in space plasmas, and such turbulence has a significant impact on the large-scale properties of the solar wind flow \citep{UsmanovEA09,KryukovEA12,WiengartenEA15,ShiotaEA17}.
The solar wind also provides a natural laboratory for {\it in situ} studies of the nonlinear dynamics of large-amplitude waves and turbulence \citep{Barnes79a,BrunoCarbone13}.
Here we will  examine statistical properties of
spatial regions, or domains, 
that have nearly constant magnetic field magnitude,
using {\it Parker Solar Probe} datasets.  
As we shall presently elaborate upon,
these domains 
have special relationships both to Alfv\'enic turbulence 
and to dynamical plasma relaxation processes.
Consequently 
the statistics of these domains of nearly constant field magnitude
may play an important role in understanding the 
nature and dynamics of the solar wind.

Using observations of the solar wind from the \textit{Mariner 5} mission from Earth to Venus, before and after its Venus flyby, \citet{BelcherDavis71} reported that fluctuations in magnetic and velocity fields were largely consistent with Alfv\'en modes, satisfying the Wal\'en relation \citep{Walen44}
\begin{equation}
    {\bf v}=\pm \frac{\bf b}{\sqrt{\mu_0\rho}}
    \label{eq:alf}
\end{equation}
(as expressed for isotropic plasma pressure), where ${\bf v}\equiv{\bf V}-{\bf V}_0$ and ${\bf b}\equiv{\bf B}-{\bf B}_0$, 
subtracting any   large-scale (mean) fields ${\bf V}_0$ and ${\bf B}_0$, and $\rho$ is the mass density. 
If there is a mean magnetic field ${\bf B}_0$, then fluctuations
in either pure $\pm$ state 
propagate along $\mp{\bf B}_0$ at the Alfv\'en speed, relative to the mean flow speed.
Broadband turbulence as found in the solar wind \citep{Coleman68} can also approximately satisfy this relation, in which case it is commonly termed Alfv\'enic turbulence. 
Of course, a superposition of 
both states in Eq.\ (\ref{eq:alf}) is required in incompressible MHD, 
to support active nonlinear interactions, in which case
pure propagation properties are no longer present \citep{Moffatt}. 
Turning to the compressible case, such large-amplitude fluctuations in a pure $+$ or $-$ polarization state can satisfy the MHD equations
provided that the total magnitude of the field, $|{\bf B}|=|{\bf B_0+b}|$, is uniform
\citep{GoldsteinEA74,BarnesHollweg74,Barnes79a}. 
Such states are sometimes called states of spherical polarization \citep{Barnes81}.
In the solar wind, Alfv\'enic fluctuations predominantly have the sign appropriate for outward propagation relative to the solar wind plasma.

Physically, local uniformity of $|{\bf B}|$ represents balance of the magnetic pressure $B^2/(2\mu_0)$.  
One might expect approximate magnetic pressure balance to be more likely when the magnetic pressure dominates over other types of pressure, e.g., when the plasma $\beta$ (ratio of plasma pressure to magnetic pressure) is much less than unity.
Indeed, \citet{KleinEA93} found that at times of relatively low plasma $\beta$ and high solar wind speed, the minimum variance directions of magnetic and velocity fluctuations from {\it Helios} data were better aligned with the mean magnetic field. 
\add{Similar results were} found by \citet{SmithEA06-aniso} in {\it ACE} data at 1 AU
\add{and by \citet{PineEA20-aniso}} 
\add{in {\it Voyager} 1 and 2 data for 
$r$ from about 1 to 32 au; that} is, in low $\beta$
plasma, fluctuations are more anisotropic and transverse to the mean field,
which of course promotes more constant $|{\bf B}|$.

In turbulence, approximate magnetic pressure balance may result from rapid, local relaxation processes that also favor patches of flow-field alignment \citep{ServidioEA08-depress}. 
It has also been attributed to conservation of ion kinetic energy in the reference frame of observed alpha particle motion \citep{MatteiniEA15}.
In any case, the divergence requirement on the magnetic field limits the spatial region over which $|{\bf B}|$ can remain uniform \citep{Barnes79a}.

{\it In situ} measurements by various space missions, especially the {\it Helios} missions that collected data as close as 0.29 au from the Sun during 1974-1984, indicated that the solar wind is more Alfv\'enic while
closer to the Sun \citep{BrunoEA07}.
The {\it Parker Solar Probe (PSP)} spacecraft, launched in 2018 August, approached within 0.17 au of the Sun during its first three solar encounters (hereafter referred to as E1-E3) and within 0.13 au during its fourth and fifth solar encounters (E4 and E5). 
{\it PSP} data have revealed even stronger Alfv\'enicity at lower radius \citep{DAmicisEA21,TelloniEA21}, with increasing dominance of outward-propagating compared with inward-propagating Alfv\'enic fluctuations \citep{ChenEA20}.
Note that Alfv\'enicity can be defined in terms of various characteristics; in addition to near uniformity of $|{\bf B}|$, it can be measured in terms of the cross-helicity, Alfv\'en ratio, residual energy, and alignment angle between {\bf b} and {\bf v} \citep{ParasharEA20ApJS}.
There has also been much excitement over {\it PSP} observations of large-amplitude {\bf b} and {\bf v} fluctuations, and in particular fluctuations that change the orientation of {\bf B} to the extent that the measured radial component $B_R$ temporarily reverses and the magnetic field lines make S-shaped bends known as ``switchbacks'' \citep[e.g.,][]{KasperEA19Nature,BaleEA19Nature,DudokDeWitEA20,MozerEA20}, which has given rise to a variety of explanations \citep[e.g.,][]{SquireEA20,FiskKasper20,RuffoloEA20,ZankEA20,SchwadronMcComas21,DrakeEA21}.
  
Previous work has remarked upon domains of nearly constant $|{\bf B}|$ punctuated by sharp changes in {\it PSP} data \citep{RuffoloEA20}.
They interpreted the domains of Alfv\'enic turbulence with nearly constant $|{\bf B}|$ as (possibly merged) mixing layers, which exhibit sharp boundaries as topological defects across which the dynamics have not yet balanced the magnetic pressure.
Close to the Alfv\'en critical zone, where the solar wind speed begins to exceed the Alfv\'en speed, such mixing layers, including enhanced fluctuations and switchbacks, should have a spatial distribution that is closely related to the underlying streamer/flux-tube structure of the solar corona, with angular scales related to supergranulation.
The domain boundaries typically involve major changes in the components of {\bf B}, often but not always including switchbacks.
These boundaries may also correspond to sharp jumps in the magnetic field vector as indicated by a high partial variance of increments \citep[PVI;][]{GrecoEA08}, and
the waiting time distribution between high PVI events in {\it PSP} data has been analyzed by \citet{ChhiberEA20-ApJS}.

In the present work we examine such domains of approximate magnetic pressure balance in more detail, starting with a prescription for identifying domains from time series of magnetic field measurements.  
We then analyze the distribution of domain duration $T_D$ and in particular the mean domain duration as a function of time $t$ and heliocentric distance $r$ in comparison with plasma parameters such as plasma $\beta$.  
We also investigate the filling fraction of domains
\add{as a measure of Alfv\'enicity}, and test simple frozen-in models of their aspect ratio by comparing the mean domain duration with the solar wind velocity relative to the spacecraft.

\section{Identification of Domains}
\label{sec:ident}

\subsection{Observational Data}

We 
proceed by analyzing 
publicly available data from 
the first five orbits (E1-E5) of {\it Parker Solar Probe}, from two instrument suites, FIELDS \citep{BaleEA16} and SWEAP \citep{KasperEA16}.\footnote{All data were downloaded from https://cdaweb.gsfc.nasa.gov/pub/data/psp/}  
We used Level 2 magnetic field data from FIELDS, which typically have a data rate of 299 Hz, and Level 3 plasma data from SWEAP's Faraday cup component (SPC), typically available at a cadence of 0.87 s.  
We then resampled the data to 1-s cadence. 
Components of vector fields are expressed in the standard spacecraft-centered orthogonal RTN coordinate system,
where +R is radial (antisunward), +T is tangential (toward increasing heliolongitude), and +N is normal (toward increasing heliolatitude). 

We used magnetic field data to identify domains of nearly constant $|{\bf B}|$, both magnetic field and plasma data to calculate plasma $\beta$, and plasma velocity data to examine the aspect ratio of the domains.
To compare the mean domain duration with $\log\beta$ over 1-d time intervals, we first calculated separate 1-h averages of $B^2$ and of $n_pT_p$, the proton number density times temperature, in the case that at least 10\% of the 1-s data were available; otherwise the 1-h period was considered to have insufficient data.  
(Squaring or multiplication was performed before averaging in order to reduce fluctuations, e.g., as $n$ and $T$ are sometimes anticorrelated.)  
These average quantities were then combined 
using the approximation $\beta \approx 2\beta_{p}$ to estimate $\log \beta$ for each hour (including protons and their associated electrons but not the contributions of alpha particles and minor ions), and these were averaged to determine $\langle\log\beta\rangle$ for the 1-d time period.
We should keep in mind that the plasma temperature from SPC measurements, which contributes to the calculation of plasma pressure and plasma $\beta$, is mainly derived from the measured distribution of $V_R$.
If more than half of the 1-h periods had insufficient data, then that 1-d time period was considered to have insufficient data.
 
We made use of two data products from SWEAP/SPC for the solar wind velocity: 
1) V\_MOMENT\_SC, the solar wind velocity relative to the {\it PSP} spacecraft, and
2) V\_MOMENT, the solar wind velocity relative to an inertial frame, that is, the spacecraft velocity has been subtracted out of the solar wind velocity 
measurement.\footnote{See the \href{http://sweap.cfa.harvard.edu/Data.html}{SWEAP Data User's Guide.}}
We first used a time-domain Hampel filter \citep{Davies1993Hampel}, with a filtering interval of 120 s and outliers identified as values more than three times larger than the local standard deviation.
Also, we found that these two data products have some independent and spurious fluctuations.
This provides a useful way to cross-check the data by subtracting the two velocity vectors to infer the spacecraft velocity, which must obey the laws of orbital mechanics.
From this we determined estimates of the spacecraft orbital energy and angular momentum, which should remain nearly constant (except at times of Venus encounters, which changed {\it PSP}'s orbit).
We first rejected 1-s data values for which these quantities deviated greatly (by $\gtrsim$50\%) and performed 1-h averages in the case that at least 10\% of the 1-s data were available; otherwise the 1-h period was considered to have insufficient data.
Then we examined the distributions of the 1-h averages of these quantities, fit the peak regions by a Gaussian function, and rejected 1-h time periods that deviated from the mean by more than three standard deviations.  
When computing a 1-d moving average, if more than half of the 1-h periods therein were missing or rejected, then that 1-d time period was considered to have insufficient velocity data.

\subsection{Definition and Selection of Domains}

Domains of Alfv\'enic fluctuations were conceptually introduced by \citet{RuffoloEA20}. 
Here, we have developed a definition of contiguous domains of nearly constant magnetic field magnitude $|{\bf B}|$, which can be physically considered as domains of approximate magnetic pressure balance, or Alfv\'enic domains.
Note that not all times are assigned to a domain.
Defining domains as contiguous is somewhat different from the concept of \citet{RuffoloEA20}, as we will discuss in Section 4.

\begin{figure*}[t]
\includegraphics[width=\textwidth]{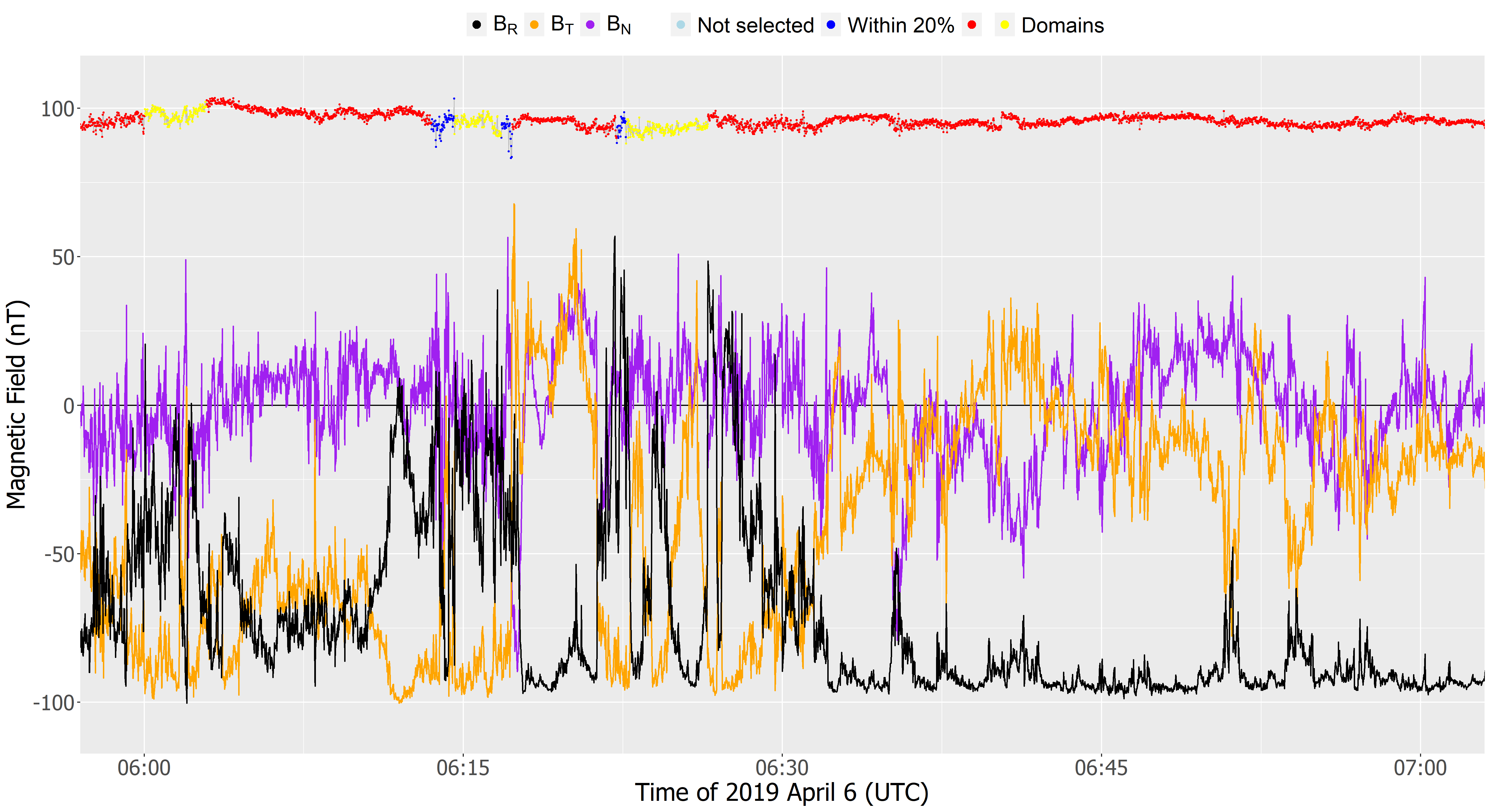}
\caption{ 
Magnetic field magnitude $|{\bf B}|$, with coloring to indicate selected domains of approximate magnetic pressure balance, and components $B_R$ (black), $B_T$ (orange), and $B_N$ (purple) from 1-s sampling of {\it PSP}/FIELDS measurements during one hour (2019 April 6, 06.00-07.00 UTC), 32 h after the second {\it PSP} perihelion. 
Along the top trace, showing values of $|{\bf B}|$, light blue points are at times that were not selected in the first stage of domain selection, dark blue is for time periods with $|{\bf B}|$ constant within 20\% but rejected in the second stage of domain selection, and alternating red and yellow coloring indicates selected domains.
Domains are interrupted by particularly strong fluctuations in $\bf B$, some but not all of which involve reversals in the sign of $B_R$, i.e., switchbacks.
}
\label{fig:B_3comp_1hr}
\end{figure*}

\begin{figure*}
\includegraphics[width=\textwidth]{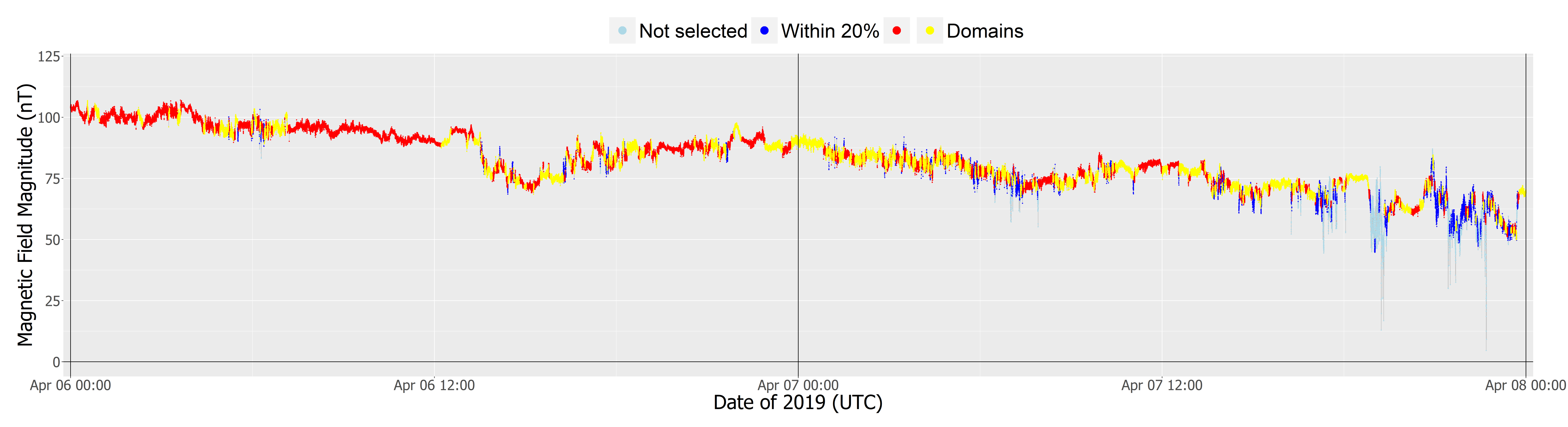}
\caption{ 
Magnetic field magnitude $|{\bf B}|$ and domains of approximate magnetic pressure balance identified from 1-s sampling of {\it PSP}/FIELDS measurements during 2019 April 6-7, a time period from 25 to 73 hours after the second {\it PSP} perihelion.
Coloring as in Figure 1.
The mean domain duration was much longer on April 6, in association with lower plasma $\beta$ on that day.
}
\label{fig:Bmag_2019Apr6-7}
\end{figure*}

Here domains are selected in two stages:
\begin{enumerate}
    \item Throughout the domain, the difference between the maximum and minimum magnetic field magnitude, $|{\bf B}|_{\text{max}}-|{\bf B}|_{\text{min}}$, should remain within a first-stage fractional tolerance of the minimum value ($f_1|{\bf B}|_{\text{min}}$).
    \item During each 600-s time period within the domain (or if the domain is shorter than 600 s, then over the entire domain), $|{\bf B}|_{\text{max}}-|{\bf B}|_{\text{min}}$ should remain within a second-stage fractional tolerance of the minimum value ($f_2|{\bf B}|_{\text{min}}$).
\end{enumerate}
We further require that domains have a duration of at least 60 s, in order to limit the number of very small domains.  
We tried changing this minimum duration, and found that the effect on the mean domain duration was roughly uniform in time and did not significantly affect trends in our analysis.  
Therefore, we use 60 s as a time that is significantly shorter than the turbulent correlation time \add{\citep[see][]{ParasharEA20ApJS,ChenEA20}} while much longer than the 1-s cadence of our dataset.
We also limit domains to contiguous time periods with no missing data. 

The purpose for using two stages is to require the entire domain to remain within a fractional tolerance $f_1$ and to require variations within that tolerance to be relatively gradual, within a smaller fractional tolerance $f_2$ over 600 s, a value chosen to be of the order of the correlation time of the 
\add{turbulence.} 
In other words, turbulent fluctuations are allowed to vary within $f_2$ while longer-time variations are allowed within a larger tolerance $f_1$. 
We chose to specify these tolerances with a 2:1 ratio, so that $f_1=2f_2$. 

To choose the tolerance values, we compared the results of domain selection for PSP data with different tolerances ranging from $f_2=0.02$ to 0.14.
We monitored the filling fraction, i.e., the fraction of 1-s data that are assigned to a domain, with the view that a useful definition of domains should assign some but not all times to Alfv\'enic domains.
Throughout the first 5 {\it PSP} orbits, we found that a high tolerance frequently yields a saturated filling fraction (FF=1) at low radius and a low tolerance frequently gives zero filling fraction at high radius.   
The intermediate values $f_2=0.08$ and 0.1 yield a filling fraction that usefully exhibits time variations (without saturation) at both low and high radius within the {\it PSP} orbits.
Among these two values, $f_2=0.1$ (which implies $f_1=0.2$) provides a clearer relation between the mean domain duration and plasma $\beta$, so we adopt these as useful tolerance values.
In summary, we define domains as contiguous time periods (of at least 60 s) when $|{\bf B}|$ remains constant to within 20\% over the entire domain (first stage of selection) and within 10\% over all 600-s time periods within the domain (second stage of selection).

As a check that the selection algorithm was implemented correctly, different members of our team developed separate computer programs to analyze the same data sets and obtained identical results, both for the numbers of domains and their durations.

\subsection{Examples and Properties of Domains}

\add{Figures 1 and 2} show 
\add{examples} of the results of domain selection using the above  prescription.
\add{Figure 1 includes} data of magnetic field components and the magnitude $|{\bf B}|$
\add{for one hour (2019 April 6, 06:00-07:00 UTC), 32 hours after the second {\it PSP} perihelion, while Figure 2 indicates domains selected during a two-day time period, 2019 April 6-7, which encompasses times from 25 to 73 hours after E2 perihelion.}
Along the trace for $|{\bf B}|$, we use coloring to indicate the results of our domain identification.  
Light blue points indicate 1-s data that were not selected even at the first stage.  
\add{(Note that for the time period shown in Figure 1, all data were selected at this stage.)}
Dark blue indicates data that were selected in the first stage, i.e., were part of a time period during which $|{\bf B}|$ was constant within 20\%, but were not selected in the second stage.  
Finally, the time periods selected as domains are indicated by alternating red and yellow coloring.  
Red and yellow colors have the same meaning, and are alternated to assist in distinguishing the boundaries of domains.

From Figures 1 and 2, 
one may observe that 
some time periods of up to several hours 
appear to have nearly constant $|{\bf B}|$, yet have alternating coloring, indicating that our algorithm selects multiple separate domains.
In such cases, the key role of the algorithm is to identify interruptions of domains due to excessively rapid variation of the magnetic field magnitude.  
In some cases the interruption is an inclusion of different $|{\bf B}|$ between similar values before and afterward; in some other cases it represents a step from one level of $|{\bf B}|$ to another.
Our present analysis does not distinguish between these possibilities, and in this sense may deviate from the conception of domains by \citet{RuffoloEA20}, who considered a domain boundary to involve a step from one level of $|{\bf B}|$ to another.

In Figure 1, we often see large fluctuations of magnetic field components within a domain of nearly constant $|{\bf B}|$.
These are likely examples of large-scale Alfv\'enic fluctuations.
Such fluctuations within a domain can have large amplitudes in any of the components, including the radial component $B_R$.
During this time period, $B_R$ was most commonly near $-|{\bf B}|\approx-100$ nT, while fluctuations are termed ``switchbacks'' if they are sufficiently extreme that $B_R$ reverses sign.  
In our view, switchbacks that occur within an Alfv\'enic domain are one type of large-amplitude Alfv\'enic fluctuation out of a continuum of possible amplitudes and directions \citep[see also][who discuss a continuum of angular deflections of {\bf B}, not only whether a deflection exceeds 90$^\circ$ so that $B_R$ reverses sign]{DudokDeWitEA20}.

Figure 1 also shows various examples of domain boundaries, where domains are interrupted.
These usually involve particularly strong fluctuations of magnetic field components, many but not all of which are switchbacks.
For example, the first transition from a ``yellow'' to ``red'' domain, from 06:02:54 to 06:02:55 UTC, was not associated with a switchback (i.e., $B_R$ did not reverse sign).
Thus domain boundaries and switchbacks are related, but not quite the same set of events: there are switchbacks that do not interrupt domains, and domain boundaries that do not include switchbacks.

Finally, we note that in Figure 2, which shows domains over 2019 April 6-7, the early part of April 6 clearly has a few domains of very long duration (up to several hours), while April 7 generally has much shorter domains.
Such time variations in the mean domain duration and distribution of domain durations, as well as their filling fraction, are the subject of further analysis in this work.




\begin{figure}[t]
    \centering
    \includegraphics[width=0.4\textwidth]{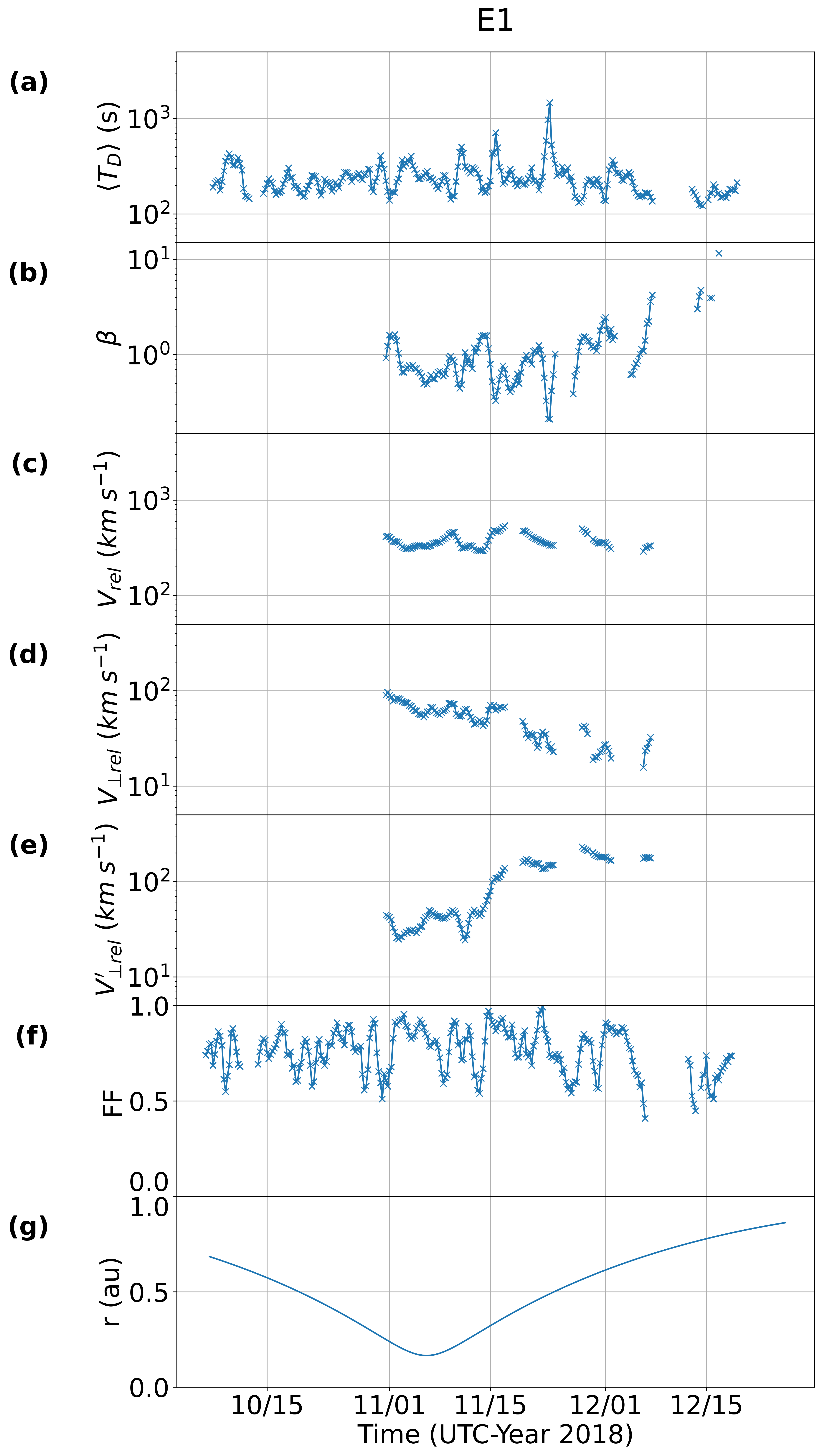}
    \label{fig:E1multiplot}
\caption{Time series during the first {\it PSP} orbit. Mean values of
(a) domain duration, $T_D$; 
(b) plasma $\beta$;
(c) magnitude of the solar wind velocity relative to the spacecraft, $V_{rel}$; 
(d) magnitude of the solar wind velocity relative to the spacecraft and perpendicular to the radial direction, $V_{\perp rel}$; 
(e) magnitude of the solar wind velocity relative to the spacecraft and perpendicular to the Parker spiral direction, $V'_{\perp rel}$; (f) filling fraction, FF;
(g) radius from the Sun, $r$.
Panels (a)-(f) show running 1-d averages with 6-h cadence, plotted at the centroid of the averaging interval.
The largest values of mean domain duration $\langle T_D\rangle$ are associated with periods of low plasma $\beta$.  
Results for $\langle T_D\rangle$ are inconsistent with inverse proportionality to $V_{\perp rel}$ or $V'_{\perp rel}$, as would be expected if domains were elongated structures along the radial or Parker spiral direction, respectively.
}
\end{figure}

\begin{figure}[t]
    \centering
    \includegraphics[width=0.4\textwidth]{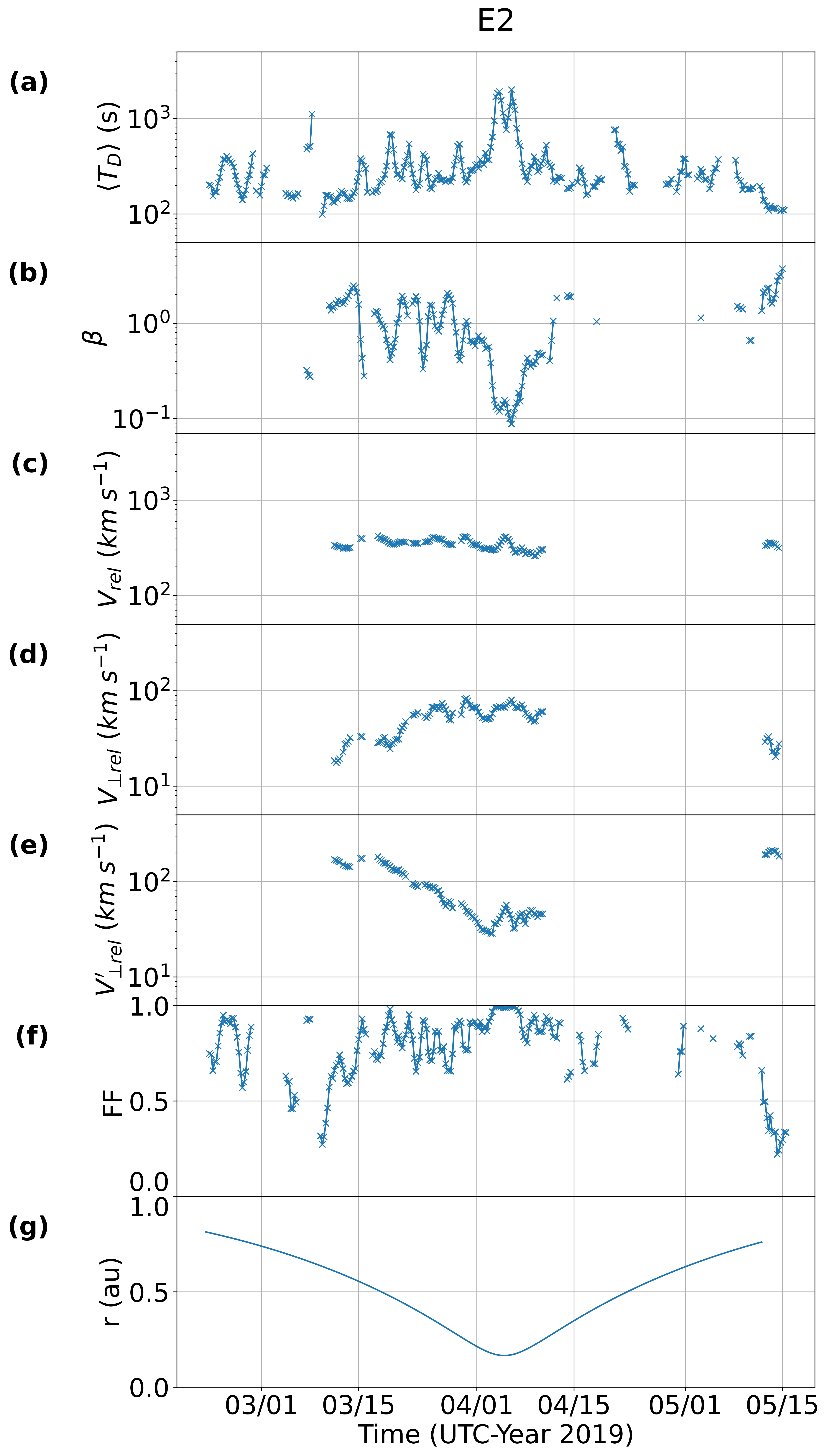}
    \label{fig:E2multiplot}
\caption{Time series during the second {\it PSP} orbit,
as in Figure 3.
The largest values of mean domain duration $\langle T_D\rangle$ are associated with periods of low plasma $\beta$.  
Results for $\langle T_D\rangle$ are inconsistent with inverse proportionality to $V_{\perp rel}$, as would be expected if domains were elongated structures along the radial direction.}
\end{figure}

\begin{figure}[t]
    \centering
    \includegraphics[width=0.4\textwidth]{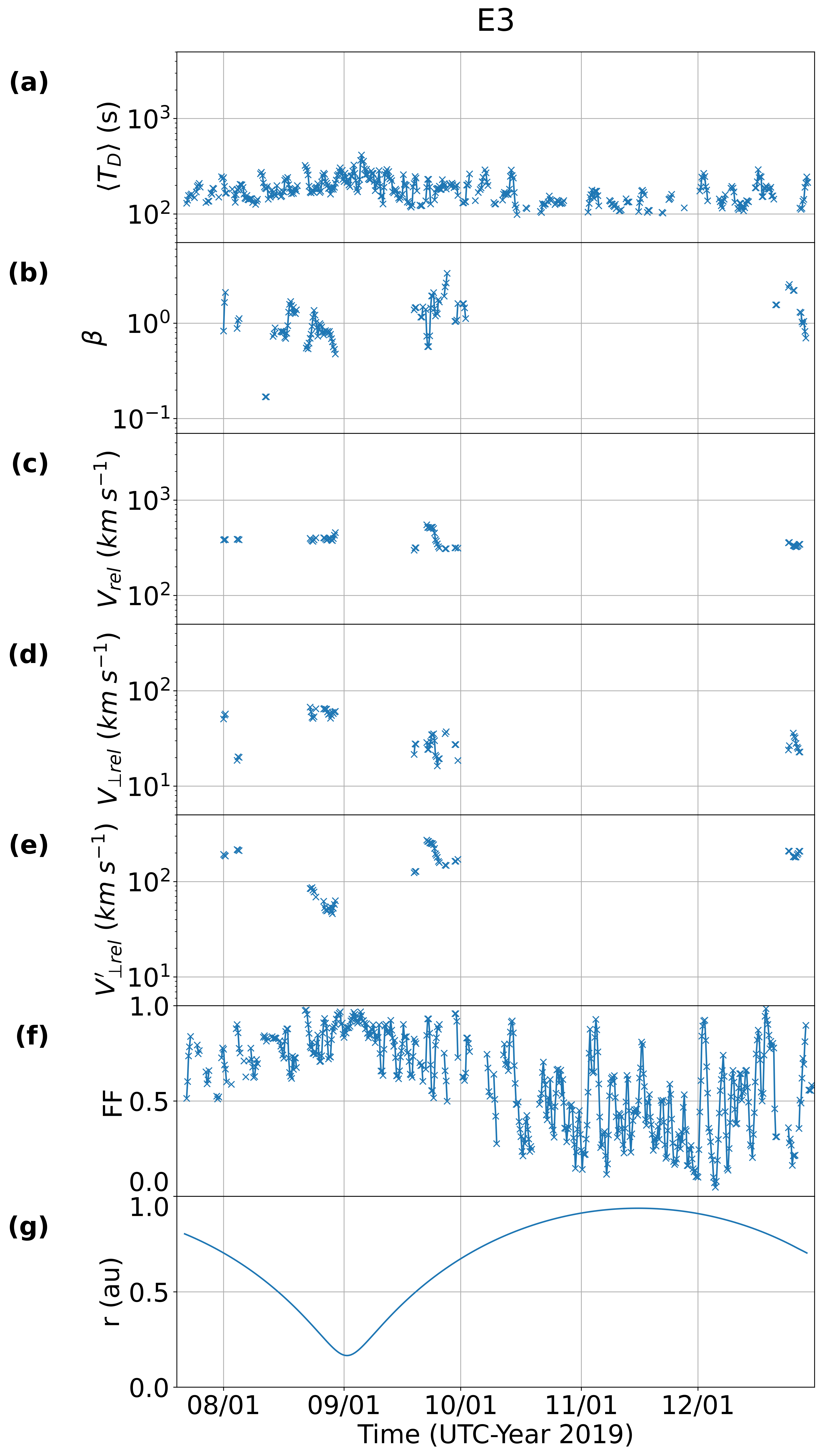}
    \label{fig:E3multiplot}
\caption{Time series during the third {\it PSP} orbit,
as in Figure 3.
Results for $\langle T_D\rangle$ are inconsistent with inverse proportionality to $V_{\perp rel}$ or $V'_{\perp rel}$, as would be expected if domains were elongated structures along the radial or Parker spiral direction, respectively.
The filling fraction, FF, has a strong inverse relation with the radius, $r$.}
\end{figure}

\begin{figure}[t]
    \centering
    \includegraphics[width=0.4\textwidth]{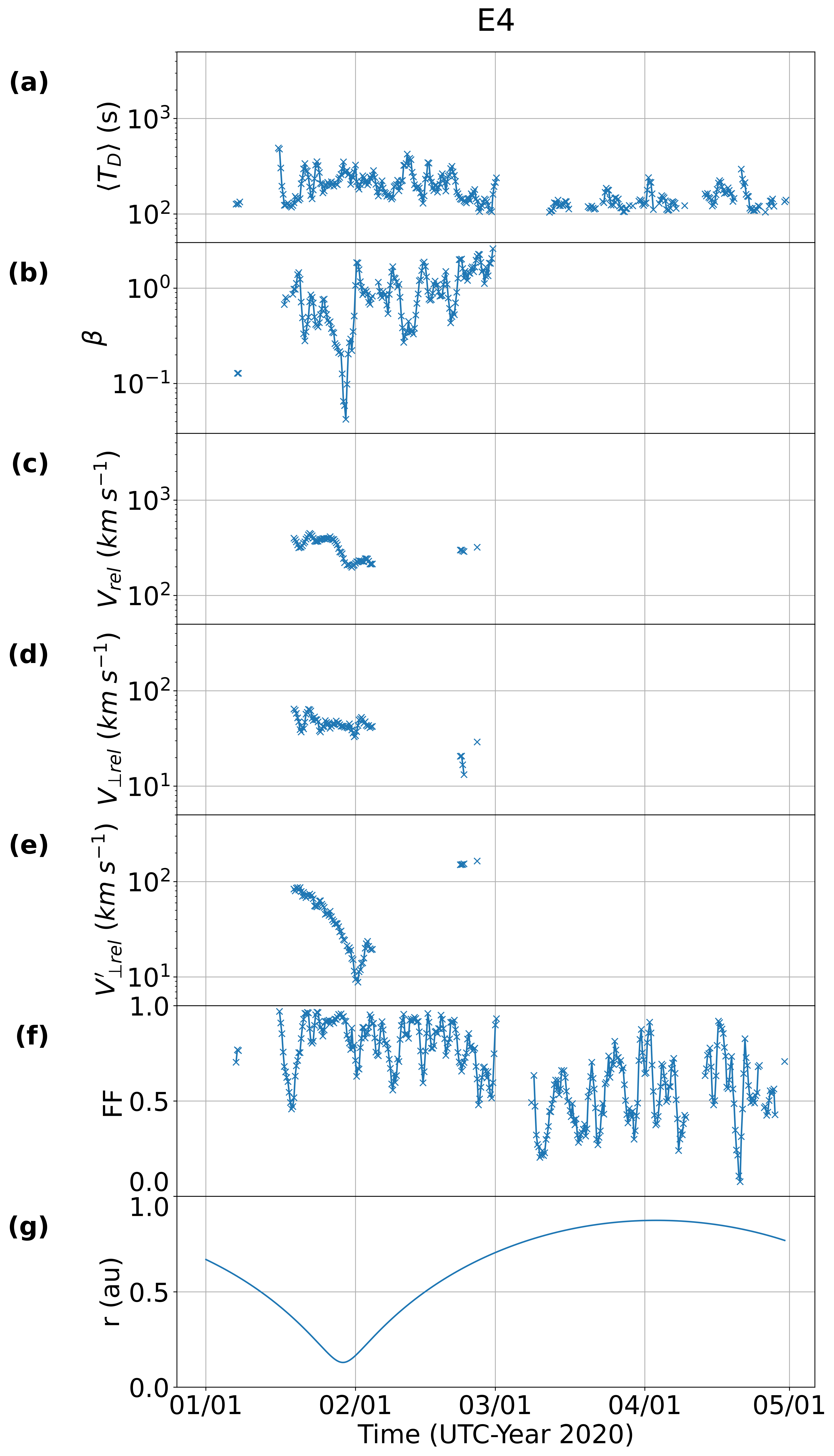}
    \label{fig:E4multiplot}
\caption{Time series during the fourth {\it PSP} orbit,
as in Figure 3.
Results for $\langle T_D\rangle$ are inconsistent with inverse proportionality to $V_{\perp rel}$ or $V'_{\perp rel}$, as would be expected if domains were elongated structures along the radial or Parker spiral direction, respectively.
The filling fraction, FF, has an inverse relation with the radius, $r$. }
\end{figure}

\begin{figure}[t]
    \centering
    \includegraphics[width=0.4\textwidth]{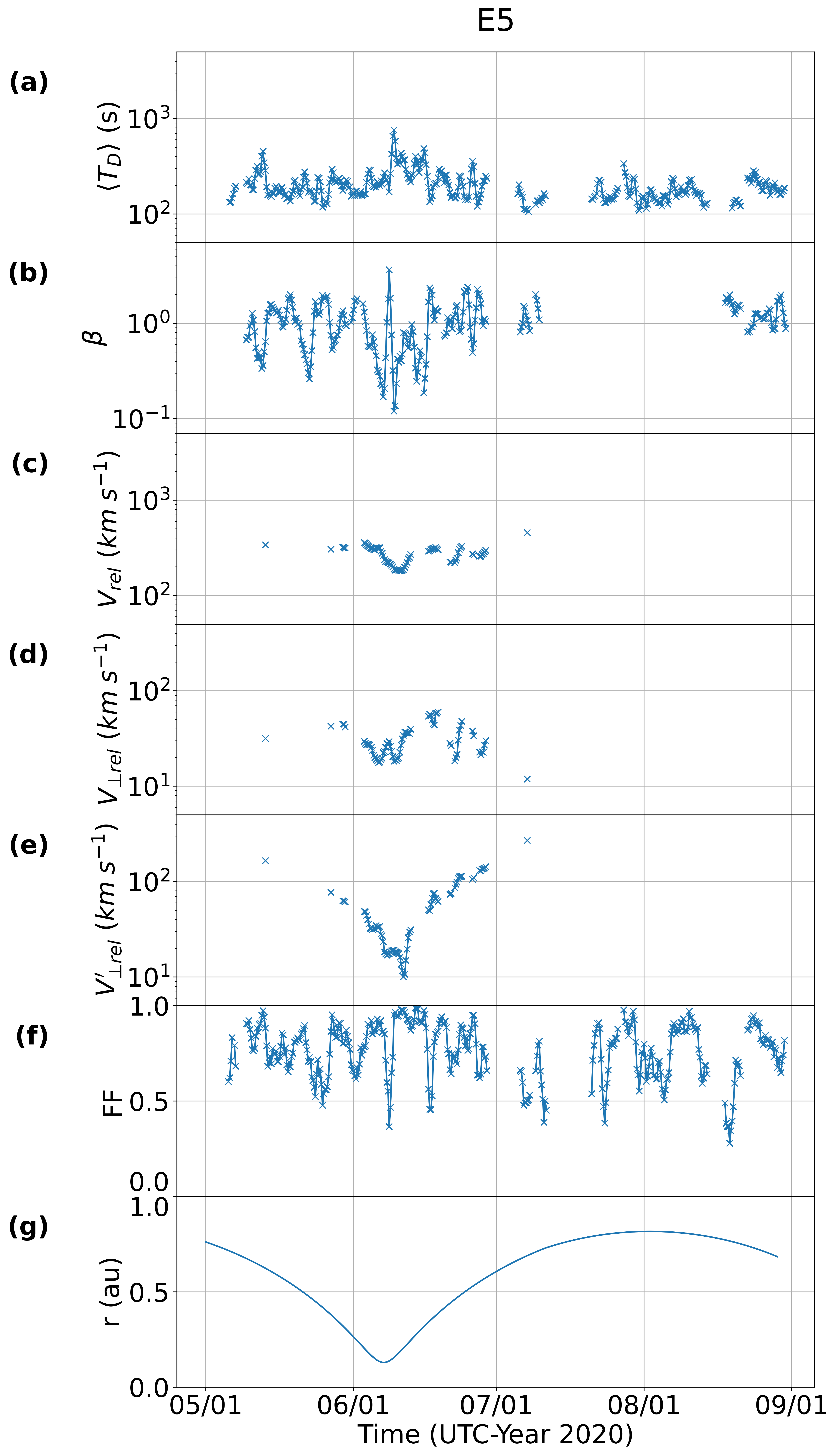}
    \label{fig:E5multiplot}
\caption{Time series during the fifth {\it PSP} orbit,
as in Figure 3.
Results for $\langle T_D\rangle$ are inconsistent with inverse proportionality to $V'_{\perp rel}$, as would be expected if domains were elongated structures along the Parker spiral direction.
}
\end{figure}

\begin{figure}
\centering
    \includegraphics[width=0.45\textwidth]{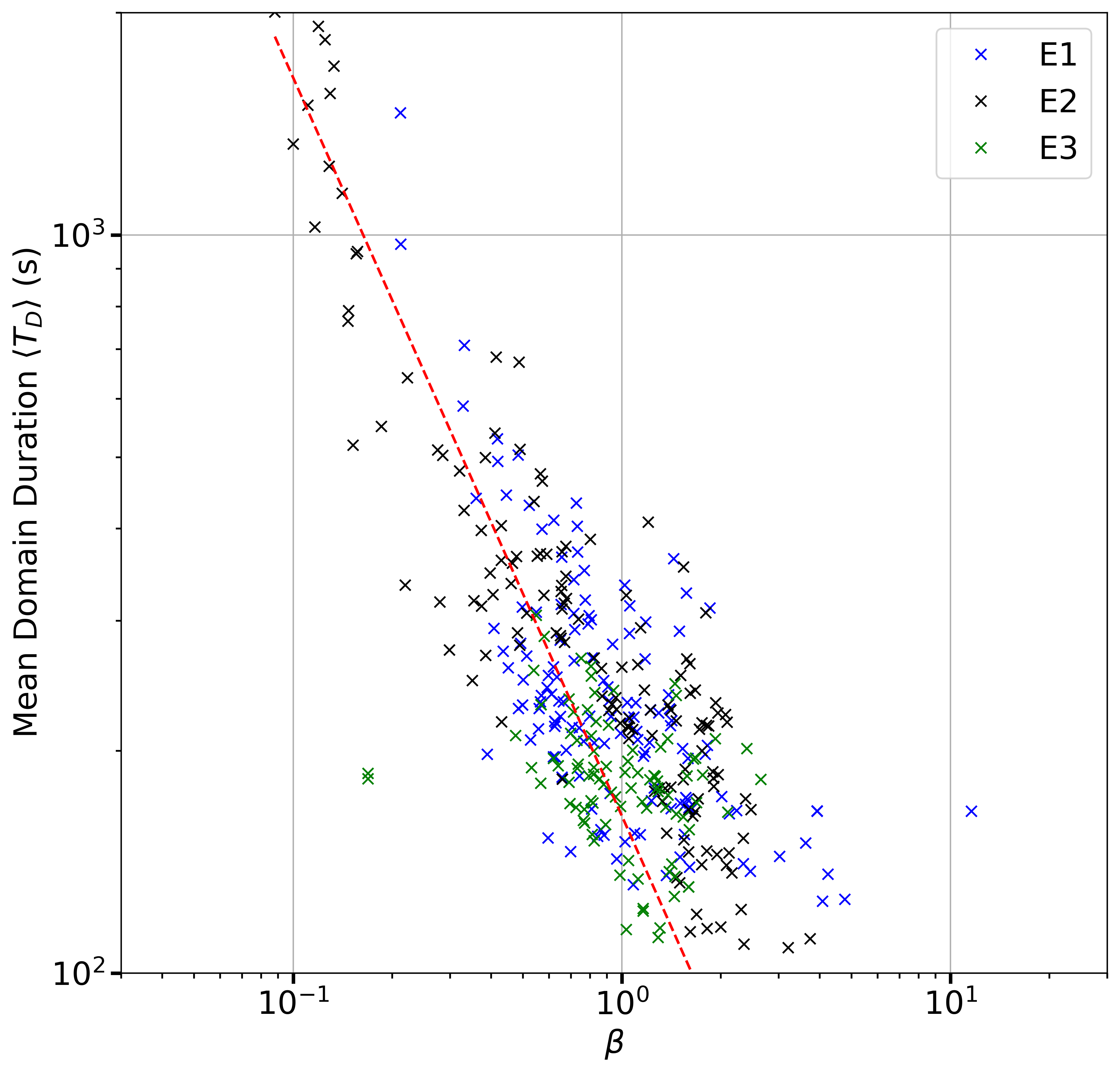}
    \includegraphics[width=0.45\textwidth]{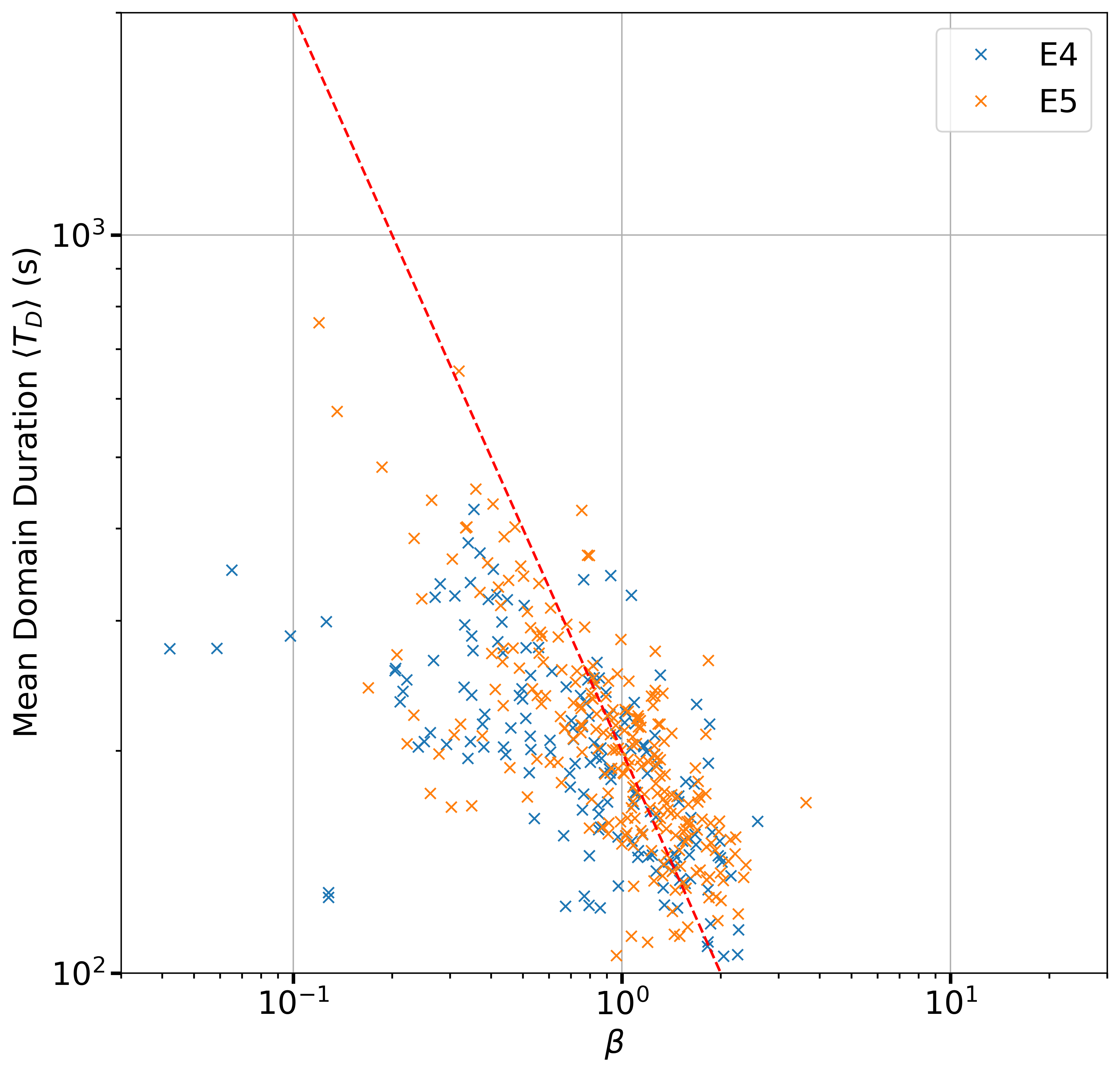}
\caption{ 
Mean domain duration $\langle T_D\rangle$ as a function of plasma $\beta$ for (a) E1-E3 and (b) E4-E5.
Each data point represents a running 1-d average with 6-h cadence.  
There is a general trend of decreasing $\langle T_D\rangle$ with increasing $\beta$.
Dashed line represents $\langle T_D\rangle\propto\beta^{-1}$.
}
\label{fig:duration-vs-beta}
\end{figure}

\section{Characterization of Domains}\label{sec:character}

We proceed to characterize domains of approximate magnetic pressure balance in data from the first five orbits of {\it PSP}.
We first examine variations in the mean domain duration. Average duration 
varies strongly with plasma $\beta$, with longer average duration associated with lower $\beta$.  
We then examine distributions of domain duration for high or low ranges of plasma $\beta$, which can be characterized by broken power laws.
Next, we consider the aspect ratio of the domains, in the context of a frozen-in approximation.
We also report the filling fraction of domains, which is strongly variable but overall has a decreasing trend with increasing distance from the Sun.

\subsection{Relationship of Domain Duration with Plasma $\beta$}

Now let us consider the variation of the mean domain duration and its dependence on plasma $\beta$.  
Given that individual domains can last for up to several hours (see Figure 2), we choose to average over 1-d time periods. 
To reduce the sensitivity to the day boundary, and to provide some information about time variation over scales less than one day, we determine the mean domain duration and other quantities over running 1-d averages computed with a 6-h cadence.

Such averages of the mean domain duration and related quantities are shown as a function of time (centroid of the averaging interval) in Figures 3-7 for the first five {\it PSP} orbits, E1-E5, respectively.
For each of these figures, panel (a) shows the mean domain duration, $\langle T_D\rangle$.
Panel (b) shows the plasma $\beta$, derived from {\it PSP} data as described in Section 2.1.
Panels (c)-(e) show measures of the solar wind velocity relative to the spacecraft, which will be discussed in Section 3.3 in connection with the 
aspect ratio of domains.
Note that for Figures 3-7, panels (a)-(e) all use logarithmic vertical scales with the same spacing per decade to facilitate quantitative comparison.
Panel (f) shows the filling fraction, and (g) plots the radial distance from the Sun.

Figures 3 and 4 show a clear inverse association between the mean domain duration and plasma $\beta$ during E1 and E2.
During these orbits, the 1-d-averaged plasma $\beta$ varied over $\sim$1.5 orders of magnitude.
During E3 (Figure 5), the relevant plasma data were available for relatively few time periods, which mostly exhibited weaker variations in $\beta$, making it difficult to identify a clear trend (see Figure 5).  
Figure 8(a) combines data from these first three orbits, each of which 
reached a perihelion distance of 0.17 au, to demonstrate 
that $\langle T_D\rangle$ generally decreases with increasing plasma $\beta$.
For comparison, the dashed line shows a trend of inverse proportionality, $\langle T_D\rangle\propto\beta^{-1}$, though with considerable scatter in the data relative to that trend.

Near E2 perihelion, there was an interesting time period with 1-d-averaged $\beta$ of $\sim$0.1 and the largest mean domain durations found in all of the first five orbits (see also Figures 1 and 2).
This time period was examined in detail by \citet{RouillardEA20}, including a comparison of {\it PSP} data with white-light coronagraph images from the LASCO instrument suite on the {\it Solar and Heliospheric Observatory} ({\it SOHO}) mission and other instruments.
They found that near E2 perihelion, {\it PSP} was usually inside streamers with high normalized density $nr^2$, but for two time periods on 2019 April 3-6 and April 6-7, {\it PSP} exited these streamers and entered regions of lower normalized density.  
These correspond to the times of very low $\beta$ and highest mean domain duration in our analysis.


\begin{figure*}
\centering
\includegraphics[width=\textwidth]{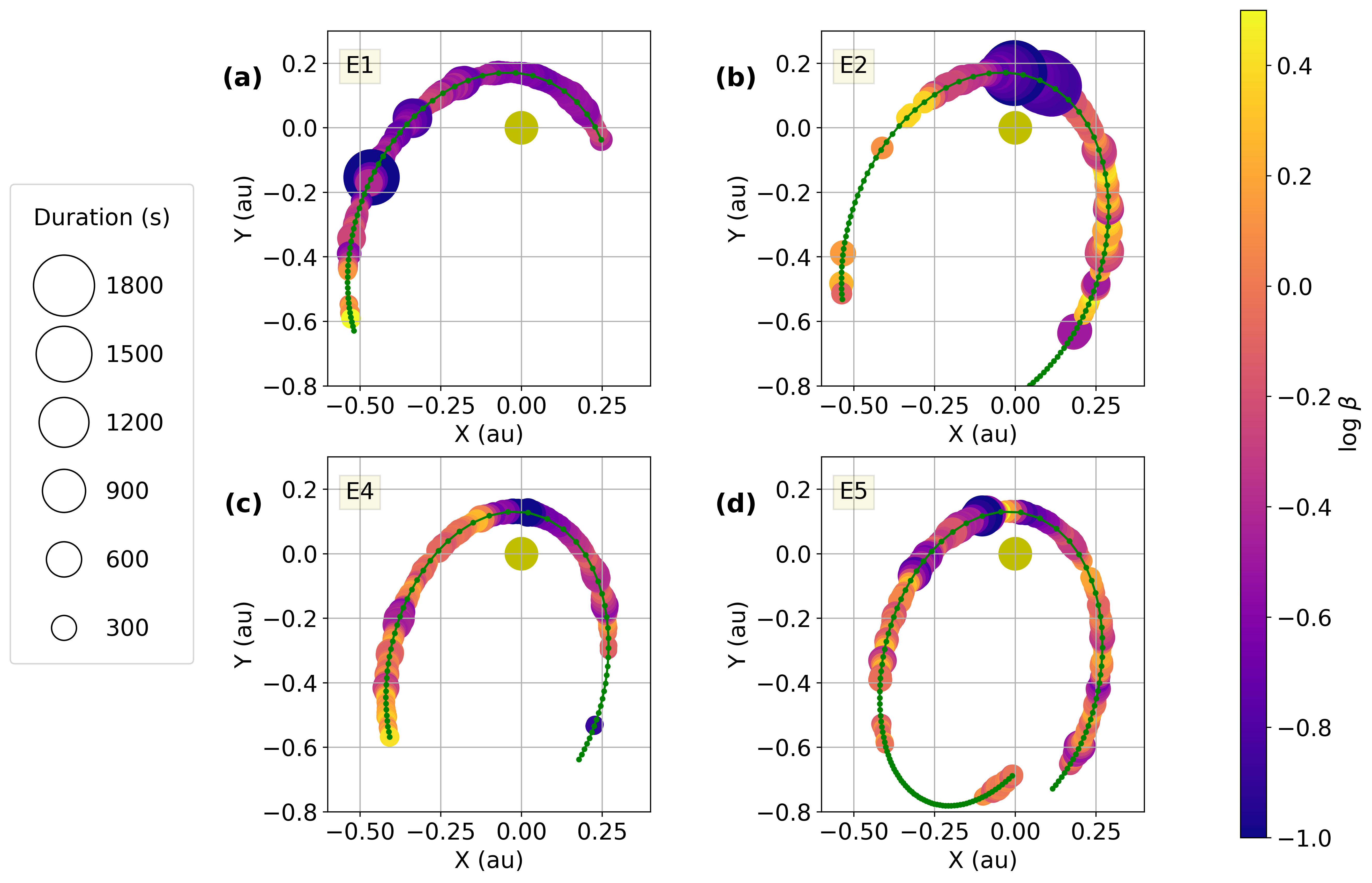}
\caption{ 
Mean duration of domains of approximate magnetic pressure balance $\langle T_D\rangle$ (indicated by circle size) and plasma $\beta$ (indicated by color scale) as a function of the location of {\it PSP} as it orbited counterclockwise around the Sun (yellow circle at $x=0$, $y=0$), for orbits E1, E2, E4, and E5.
Small green markers represent the location of {\it PSP} at the start of each day.
Data of $\langle T_D\rangle$ and $\beta$ are shown for
1-d averages with 6-h cadence.  
It is seen that $\langle T_D\rangle$ generally has an inverse relationship with $\beta$ and is otherwise not dependent on distance from the Sun. 
}
\label{fig:5orbits}
\end{figure*}

\begin{figure}
\centering
\includegraphics[width=0.45\textwidth]{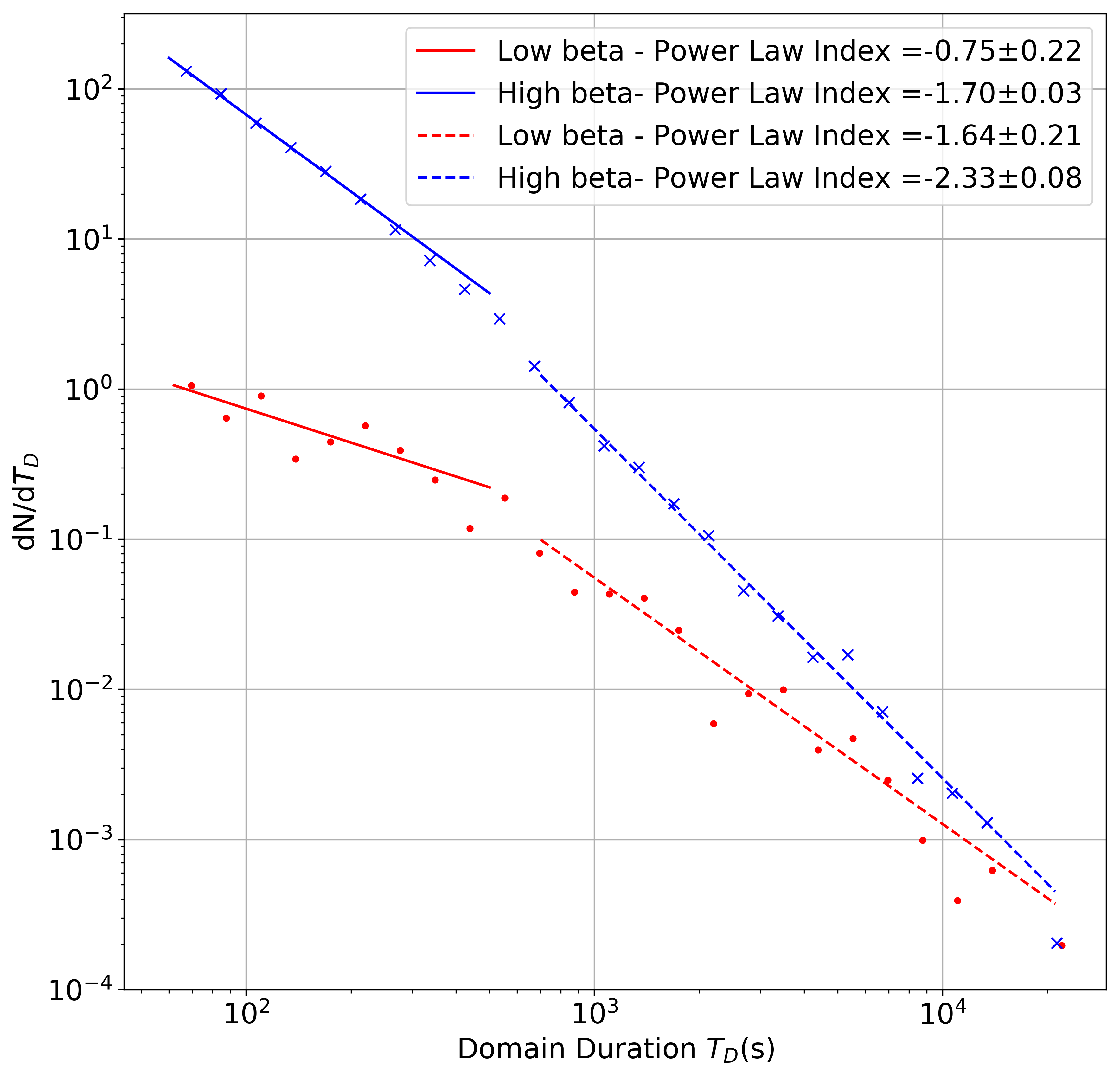}
\caption{
Distributions of domain duration $T_D$ during E2, 
for times of $\beta<0.16$ (red symbols) and $\beta>0.56$ (blue symbols).
The distributions have a broken power-law form with a break on the order of the correlation time.
The lower $\beta$ range has stiffer magnetic resistance to disturbance of magnetic pressure balance, leading to a distribution weighted toward higher $T_D$ values.
}
\label{fig:histE2highlowbeta}
\end{figure}

For encounters 
E4 and E5, plasma $\beta$ data again exhibit strong variations with time, as illustrated in 
Figures 6 and 7.
The mean domain duration generally exhibits an inverse association with $\beta$.
There are some notable exceptions to this, 
including some time periods near or shortly before E4 and E5 perihelia, when $\beta$ had very low values while $\langle T_D\rangle$ was only moderately high.
Figure 8(b) combines data for $\langle T_D\rangle$ as a function of $\beta$ for E4 and E5, when {\it PSP} reached a perihelion 
at 0.13 au.
For the E4 and E5 orbits, there is still an overall inverse relationship, though it is not as dramatic as for E1-E3. 

Low $\beta$ indicates
a dominance 
of magnetic pressure over  plasma pressure, 
a condition that may favor maintenance of 
good pressure balance over longer time periods. 
Therefore, it is physically reasonable that low $\beta$ is associated with longer mean domain duration $\langle T_D\rangle$.  
The results in Figure 8 show a continuous and extended relationship, in which increasing $\beta$ is associated with decreasing $\langle T_D\rangle$, even for relatively high $\beta>1$.
As mentioned in the introduction, previous work also found that at times of relatively low plasma $\beta$, the minimum variance directions of magnetic and velocity fluctuations were better aligned with the mean magnetic field,
consistent with relative constancy 
of $|{\bf B}|$ \add{\citep{KleinEA93,SmithEA06-aniso,PineEA20-aniso}.} We address this issue further in the discussion section. 

Figure \ref{fig:5orbits} displays the mean domain duration $\langle T_D \rangle$ and plasma $\beta$ as a function of {\it PSP}'s location along its orbit for E1, E2, E4, and E5.  
Note that these represent views from ecliptic North, in which {\it PSP} orbits counterclockwise around the Sun (yellow circle at $x=0$, $y=0$).
We do not show a plot for E3 because of the relative paucity of plasma data. 
The circle size indicates the domain duration, and the color scale indicates log($\beta$). 
During most orbits, $\beta$ was lowest near each encounter's perihelion, though this was not the case for E1, during which $\beta$ was lowest over two weeks after the perihelion. 

Figures 3-7 and Figure 9 indicate no apparent relationship between the mean domain duration $\langle T_D\rangle$ and distance $r$ from the Sun, except insofar as $\langle T_D\rangle$ depends on plasma $\beta$.


\subsection{Domain Duration Distribution}
Because plasma $\beta$ has a strong effect on the mean domain duration, as seen above, we have also examined its effect on the distribution of the domain duration $T_D$.
Such distributions are shown in Figure \ref{fig:histE2highlowbeta} for times of relatively low or high plasma $\beta$ during the second orbit, E2. 
Specifically, the low $\beta$ range is for $\log(\beta) < -0.8$ ($\beta<0.16$, red symbols).  
In this case, $\beta\ll1$ and the magnetic pressure is dominant over plasma pressure.
The relatively high $\beta$ range is for $\log(\beta) > -0.25$ ($\beta>0.56$, blue symbols), which for this data set corresponds to $\beta\sim1$ (see Figure 4), with roughly equipartitioned magnetic and plasma pressure. 
These distributions were computed by acquiring 
counts in bins of $T_D$ values, each bin
having 
equal width on a log scale.
We then plot the distribution function $dN/dT_D$, defined as the number of domains in each bin divided by the $T_D$ bin width. 
In this way the 
distributions are normalized to the number of cases.
Therefore, for higher $\beta$, the distributions have greater $dN/dT_D$ simply because there were more instances of 
higher $\beta$ in the data set. 

We find the distributions to have a broken power-law form, with a break at $ T_{D,b}  \approx 500$ s, which is on the order of the correlation time
\add{\citep[noting that the correlation time varies significantly over the initial {\it PSP} orbits;][]{ChenEA20}}.  
We have performed separate power-law fits, $dN/dT_D\propto T_D^\alpha$, to each distribution for $T_D\ll T_{D,b}$ and $T_D\gg T_{D,b}$.
For the lower $\beta$ range we find power-law indices $\alpha=-0.75\pm0.22$ for low $T_D$ and 
$\alpha=-1.64\pm0.21$ for high $T_D$.
For the higher $\beta$ range we find power-law indices $\alpha=-1.70\pm0.03$ for low $T_D$ and 
$\alpha=-2.33\pm0.08$ for high $T_D$.

In the previous subsection, we reported that times of lower $\beta$ generally have higher values of $\langle T_D\rangle$, which is physically reasonable because they have stiffer magnetic resistance to disturbance of magnetic pressure balance.
Figure \ref{fig:histE2highlowbeta} and the power-law indices demonstrate that at lower plasma $\beta$, not only is the mean duration greater, but the entire distribution is quite different, with higher values of the power-law indices and a relative enhancement of longer domain durations.
We will discuss the possible physical significance of the broken power-law distributions in Section 4.

\subsection{Aspect Ratio}



We now consider hypotheses that our data for the mean domain duration $\langle T_D\rangle$ can be described by spatial structures of various aspect ratios that convect past the spacecraft along with the solar wind.
Such convection would imply that $\langle T_D\rangle$ is inversely proportional to a measure of the relative velocity between the solar wind and the spacecraft, so we can use such a relation to test each hypothesis.
Because plasma $\beta$ has a strong influence on $\langle T_D\rangle$, we compare $\langle T_D\rangle$ with a measure of relative velocity using
time periods that have similar $\beta$ values.

First, consider the hypothesis in which domains are spherical or isotropic with a constant size distribution that does not depend on time or distance from the Sun.  
The time duration over which PSP crosses a sphere should relate to the relative speed between the solar wind and the spacecraft, i.e., $V_{rel}=|{\bf V}-{\bf V}_s|$, where {\bf V} is the solar wind velocity and ${\bf V}_s$ is the velocity of the spacecraft.  
More precisely, for a sphere of radius $a$, a random transept through the sphere has a mean length of $(4/3)a$.  
Accordingly we expect the mean duration of transit through the sphere to be
\begin{equation}
    \langle T_s\rangle = \frac{4}{3} \frac{a}{V_{rel}},
    \label{eq:sphere}
\end{equation}
where
\begin{equation}
    V_{rel}\equiv\sqrt{(V_R-V_{s,R})^2
    +(V_T-V_{s,T})^2
    +(V_N-V_{s,N})^2},
\end{equation}
and we have expressed $V_{rel}=|{\bf V}-{\bf V}_s|$ in terms of the velocity components. 
More generally, for a uniform and constant distribution of the size $a$, we can expect that $\langle T_D\rangle\propto 1/V_{rel}$. 

If domains are delineated by strong magnetic fluctuations that interrupt the near constancy of $|{\bf B}|$,  
these need not form a continuous boundary around the domain, and the actual duration of domains would depend on how frequently the spacecraft trajectory encounters such interruptions.
Thus the hypothesis for isotropic domains, and $\langle T_D\rangle\propto 1/V_{rel}$, would apply to a case where strong fluctuations that interrupt domains have a uniformly random distribution in space, or are otherwise distributed such that a trajectory along any orientation encounters such interruptions with the same spatial distribution. 



Next, suppose that domains are shaped as tubes that are highly elongated along the $R$ (radial) direction, with circular dimension $b$ in the $T$ and $N$ directions.  
This hypothesis is motivated by the work of \citet{HorburyEA20}, who found that Alfv\'enic ``spikes,'' including switchbacks, observed in {\it PSP} data have a large aspect ratio, shaped as long, thin structures oriented roughly along the radial direction. 
In this case, entering and exiting the domain is determined only in the $T$ and 
 $N$ directions, so we consider the relative velocity between the solar wind and the spacecraft in only these directions, i.e.,
 perpendicular to the radial direction.  
 A random transept in two dimensions through the circular cross-section has mean length $(\pi/2)b$.  
 Thus the mean time to cross the tube is 
 \begin{equation}
    \langle T_t\rangle = \frac{\pi}{2} \frac{b}{V_{\perp rel}}
    \label{eq:tube}
\end{equation}
where
\begin{equation}
V_{\perp rel}\equiv\sqrt{(V_T-V_{s,T})^2
    +(V_N-V_{s,N})^2}.
\end{equation}
For a distribution of the size $b$ that is uniform and constant in space, this hypothesis implies that $\langle T_D\rangle\propto1/V_{\perp rel}$.  
This hypothesis also includes the case in which domains are interrupted by strong magnetic fluctuations, if those are arranged in a roughly two-dimensional pattern with a spacing in the $T$ and $N$ directions that is statistically homogeneous, constant, and axisymmetric.  
A pattern of radially aligned magnetic flux tubes, each with a distinct value of $|{\bf B}|$, is also included in this case.

Finally, we consider a third hypothesis, similar to the second but in which domains represent structures elongated along the local Parker spiral direction, i.e., the expected direction of the large-scale interplanetary magnetic field based on the measured local solar wind speed and the rate of solar rotation \citep{Parker58}.
This hypothesis is inspired by the work of \citet{LakerEA21}, who inferred that switchbacks in {\it PSP} data are long, thin structures oriented along the Parker spiral direction.
For this hypothesis, we expect that $\langle T_D\rangle\propto1/V'_{\perp rel}$, where $V'_{\perp rel}$ is the magnitude of the projection onto a plane perpendicular to the local Parker spiral direction of the solar wind velocity relative to the spacecraft.


Therefore, we consider three measures of the relative velocity, $V_{rel}$, $V_{\perp rel}$, and $V'_{\perp rel}$, as relevant for examining these three hypotheses for extreme cases of the aspect ratio of domains. 
In Figures 3-8, these three measures are plotted in panels (c)-(e), respectively.
We then compare the time series of $\langle T_D\rangle$ (panel (a)) with those in panels (c)-(e), keeping in mind that each hypothesis implies an inverse proportionality between $\langle T_D\rangle$ and the relevant measure of the relative velocity, for times of similar plasma $\beta$.
Note that panels (a)-(e) are all plotted with the same logarithmic scale size to assist in the comparison.

For each of the first five orbits, $V_{rel}$ (panel (c)) exhibits only moderate variation, so the data do not rule out the relationship $\langle T_D\rangle\propto V_{rel}^{-1}$, as expected for spherical or isotropic domains, nor do they provide supporting evidence for such a relationship.

In contrast, for E1-E4 (Figures 3-6), the time series for $\langle T_D\rangle$ (panel (a)) and $V_{\perp rel}$ (panel (d)) are inconsistent with an inverse proportionality $\langle T_D\rangle \propto V_{\perp rel}^{-1}$.  
Similarly, for E1 (Figure 3) and E3-E5 (Figures 5-7), the time series in panel (a) and for $V'_{\perp rel}$ (panel (e)) are inconsistent with an inverse proportionality $\langle T_D\rangle \propto (V'_{\perp rel})^{-1}$.
Therefore, our results are inconsistent with domains delineated by very elongated shapes along the radial or local Parker spiral direction that convect past the spacecraft with the solar wind.

\subsection{Filling Fraction}

We use the term ``filling fraction'' (FF) to refer to the fraction of 1-s data points that are assigned to any domain, using the definition and procedure
adopted in Section 2.2.
The time series of FF for E1-E5 are plotted in panel (f) of Figures 3-7.  
Though FF exhibits strong variations, it is clear, especially for E3 and E4 (Figures 5-6), that FF is usually close to 1 near perihelion and can be significantly lower at greater distance $r$ from the Sun (plotted in panel (g)).
We have not found an association of FF with plasma $\beta$, crossing of the heliospheric current sheet, or other physical features of the solar wind, other than the distance from the Sun.

We have computed the median FF values from all five orbits for different ranges of $r$, and we found the very different median values of 0.90 for $0.1<r<0.2$ au and 0.38 for $0.9\leq r<1.0$ au.
While a high FF indicates that most of the selected data points are assigned to domains, these could either be a large number of short domains or a smaller number of longer domains, as indicated by the mean domain duration. 
The filling fraction of a time range can be interpreted as a quantitative measure of its Alfv\'enicity, so our finding that FF varies strongly with $r$ is consistent with numerous findings in the literature that the solar wind is more frequently Alfv\'enic at closer distances to the Sun (see Section 1).
 

\section{Discussion}\label{sec:disc}

{\it Parker Solar Probe} continues to explore the inner solar wind at 
previously unattained proximity to the
Sun, affording 
observers and theorists alike opportunities to 
examine distinctive features of the plasma and fields approaching 
the Alfv\'en critical region \citep{DeForestEA16,KasperEA19Nature,BaleEA19Nature}.
Much attention has been given to 
switchbacks, jets and large-amplitude, high cross-helicity
fluctuations observed 
in this region with vigorous ongoing debate 
concerning the origin and expected evolution 
of such structures \citep{SquireEA20,FiskKasper20,RuffoloEA20,ZankEA20,SchwadronMcComas21,DrakeEA21}. 
So far, 
considerable emphasis has been placed on 
quantifying Alfv\'enicity and magnetic field 
reversals (switchbacks),
as well as details
such as 
directionality of strahl and energetic particles, and general turbulence properties in the vicinity of these distinctive fluctuations. 
Here we have examined in some detail a related but 
distinct feature of the 
magnetic field near PSP perihelia that, as far as we are aware, 
has not as yet been studied and quantified 
in detail as a separate phenomenon: the 
organization of the magnetic field into local domains 
of near constant field magnitude. 

The present work defines a domain of approximate magnetic pressure balance, or Alfv\'enic domain, as a contiguous set of data over which $|{\bf B}|$ varies within prescribed limits (see Section 2.2).  
This local definition allows examination of:
\begin{enumerate}
    \item Well-defined, quantitative measures.
    \item The mean duration, which relates to plasma $\beta$ and the frequency of interruptions, which in turn relates to PVI events and switchbacks.
    \item The filling fraction (because not all times are assigned to a domain), which serves as another measure of the Alfv\'enicity of an extended time period.
    \item The aspect ratios of uninterrupted regions of nearly constant $|{\bf B}|$.
\end{enumerate}

The 
\add{domain duration} defined in this work 
\add{does} not characterize the overall constancy of $|{\bf B}|$ over an extended time period, unless that time period is free of interruptions to near-constancy of $|{\bf B}|$.
For example, consider Figures 3 and 4 of \citet{TelloniEA21}, who showed time series of the magnetic field components and magnitude for a parcel of solar wind as it passed {\it PSP} at $r\approx0.1$ au (near E6 perihelion) and later passed the Solar Orbiter (SolO) spacecraft at $r\approx1.0$ au.
They commented that the solar wind plasma was more Alfv\'enic at 0.1 au, and from their figures the time series of $|{\bf B}|$ at 0.1 au was clearly much more constant (in fractional terms) than $|{\bf B}|$ at 1.0 au.
Yet the present work finds that the mean domain duration generally does not depend on $r$, except insofar as it depends on plasma $\beta$.  
(Plasma $\beta$ can evolve with distance from the Sun.
Such evolution was not measured for the solar wind parcel studied by \citet{TelloniEA21} because SolO plasma measurements were not available for the relevant time period.)
The reason is that our domain analysis does not address the overall constancy of $|{\bf B}|$ over such extended time periods; instead, it relates to local magnetic pressure balance and turbulent relaxation.
Indeed, from Figure 3 of \citet{TelloniEA21}, it is seen that when the parcel of solar wind was at 0.1 au, there were numerous interruptions to near constancy of $|{\bf B}|$.
Such interruptions or domain boundaries may be interpreted as fault lines across which the solar wind plasma can evolve independently and thereafter have different values of $|{\bf B}|$ in different domains.
We further note that our filling fraction of domains does provide a measure of Alfv\'enicity of extended time periods, and this indeed does exhibit a strong dependence on $r$.

Domains of nearly constant $|{\bf B}|$ were introduced by \citet{RuffoloEA20}, although they were not clearly defined or systematically examined.  
That work noted an extended period of nearly constant $|{\bf B}|$ on 2018 Nov 11, 5 days after E1 perihelion, based on which they discussed the possibility that domain durations increase with increasing distance from the Sun.
(Their Figure 11(c) shows a histogram, said to be for 0000-0800 UT, but upon closer examination their data set ended at about 0325 UT.)
For this case, our domain analysis does find an uninterrupted domain for nearly 5 hours, and this was a time period of moderately low $\beta$.
More generally we find that the mean domain duration has a strong inverse relation with $\beta$, but otherwise no apparent dependence on $r$.


Given that domain boundaries represent interruptions of near constancy of $|{\bf B}|$, they are often related to strong magnetic fluctuation events, such as switchbacks and magnetic PVI events. 
For the example shown in Figure 1, most of the domain boundaries involve switchbacks, while roughly half of the switchbacks or switchback patches disrupt a domain, i.e., involve a substantial change in $|{\bf B}|$.
Therefore, our distributions of domain duration (see Figure 10), which have broken power-law forms, can be interpreted as waiting time distributions between such interruptions.
Previous work has addressed the waiting-time distributions of {\it PSP} observations of switchbacks \citep{DudokDeWitEA20} and magnetic PVI events \citep{ChhiberEA20-ApJS}. 
For both cases, the reported distributions exhibited a power-law form with a break at $\sim10^2-10^3$ s, as in our results, roughly corresponding to the turbulence correlation time.
Our results extend to durations over 10$^4$ s, allowing us to identify another power-law for $T_D>T_{D,b}$, above the break duration.  
The switchback waiting time distributions of \citet{DudokDeWitEA20} are also consistent with broken power laws,  
while the PVI waiting time distributions of \citet{ChhiberEA20-ApJS} did not extend to such long times, making it difficult to unambiguously determine the distribution shape above the break time.

For \add{short durations with $T_D<T_{D,b}$, below} the turbulence correlation scale, i.e., in the inertial range of turbulence, it is unsurprising to obtain power-law distributions as these are characteristic of scale-invariant phenomena.
For the power-law index at lower times, \citet{DudokDeWitEA20} obtained $\alpha=-1.4$ to $-1.6$, while \citet{ChhiberEA20-ApJS} obtained $\alpha=-0.6$ to $-1.3$. 
Our distributions for E2 yielded $\alpha=-0.75\pm0.22$ for $\beta<0.16$ and $\alpha=-1.70\pm0.03$ for $\beta>0.56$, which are reasonably consistent with the previous results.
\citet{ChhiberEA20-ApJS} pointed out that for comparison, the waiting time distribution of the Cantor set, a classic example of a highly clustered set of events \citep{Cantor1883}, is a power law with $\alpha=-1.63$, and generalized Cantor sets with even stronger clustering have $\alpha\to-1$.
In this sense, the $\alpha$ value that we obtain for the higher $\beta$ range (which is more common in {\it PSP} observations) already represents strong clustering similar to that of the classic Cantor set, while the lower $\beta$ range has an $\alpha$ value that represents even more extreme clustering of domain boundaries.

In our work, the domain duration distributions at $T_D>T_{D,b}$, above the break duration, also exhibit a power-law form.  
This indicates scale invariance in the occurrence of domain boundaries, even for time scales longer than the outer scale of the turbulence.
We speculate that this could be associated with the scale-invariant $1/f$ noise in the solar wind at low frequency $f$, which is believed to be driven by solar phenomena
\citep{MattGold86}, 
originating in the lower corona \citep{BemporadEA08}
or possibly 
deeper in the photosphere or below \citep{MatthaeusEA07}. 



It is not surprising that plasma $\beta$ has a strong effect on domains of approximate magnetic pressure balance, with a qualitative difference between solar wind of $\beta<1$, for which magnetic pressure dominates over plasma pressure, and $\beta\approx1$, with approximate equipartition between magnetic and plasma energy density and pressure.
Previous work has classified the slow solar wind as having two states with different ranges of $\beta$ \citep[e.g.,][]{GritonEA21,DAmicisEA21}.
However, we do find it surprising that our results indicate a quantitative, continuous relationship between the mean domain duration $\langle T_D\rangle$ and $\beta$ (see Figure 8), rather than a step or switching function between two extreme states.
Specifically, $\beta$ has a continuous effect on $\langle T_D\rangle$ values over the range $0.1<\beta<2$. 
The deviation from the overall quantitative trend for some time periods at very low $\beta$ near E4 and E5 perihelia, during which {\it PSP} was also near the heliospheric current sheet, invites further investigation.

In addition to the inverse association with $\beta$, our results are consistent with the relationship $\langle T_D\rangle\propto 1/V_{rel}$ as expected if the domain duration represents the time to traverse spatial regions, either between boundary surfaces or between a pattern of features that interrupt near constancy of $|{\bf B}|$, that are statistically homogeneous and isotropic in shape, and convect past the spacecraft with the solar wind.  
(The evidence does not specifically indicate such proportionality, but does not rule it out.)

\add{However, the} results are inconsistent with the relationships $\langle T_D\rangle\propto 1/V_{\perp rel}$ and $\langle T_D\rangle\propto 1/V'_{\perp rel}$ as expected for convection of such spatial structures if they are highly elongated along the radial or local Parker spiral direction, respectively.
This rules out, for example, the interpretation of domain boundaries as solid walls of flux tubes oriented in one of these directions.
This does not necessarily conflict with the recent observation that switchback regions are elongated along the Parker spiral direction \citep{LakerEA21}.  
Given that the domain boundaries often include switchbacks, the domains themselves may represent the space between switchbacks, in which case the aspect ratio of domains relates to the spatial patterning of switchback positions rather than the shapes of individual switchback regions.

\add{Indeed,} even if coherent structures such as current sheets are believed to relate to a flux tube structure in the solar wind, as suggested many times in the literature \citep[e.g.,][]{BartleyEA66,BrunoEA01,Borovsky08}, 2D MHD simulations of magnetic turbulence have indicated that flux structures do not have well-defined walls densely populated with current sheets; rather, current sheets are sparsely distributed at different nested flux structures \citep{SeripienlertEA10}.
We speculate that in three dimensions, the locations of current sheets may be sparsely distributed along the flux tube direction as well.
This could relieve the tension between observations of flux-tube structuring in the solar wind and the result that domains, related to the spaces between strong coherent magnetic structures, do not exhibit a large aspect ratio along the radial or Parker spiral direction.

Finally, we note that our inference that domains do not represent elongated structures in one of those directions, 
\add{and could have an isotropic shape,}
is consistent with the observation from white-light images that solar wind density fluctuations transition from radial ``striae'' near the Sun to ``flocculae'' that are nearly isotropic in the sky plane \citep{DeForestEA16}.
\add{This transition} can be attributed to shear-driven dynamics and nonlinear Kelvin-Helmholtz instabilities near the Alfv\'en critical zone that energize solar wind turbulence and lead to mixing layer dynamics that tend to isotropize the distribution of coherent structures in the solar wind \citep{RuffoloEA20}.
\add{Note that the data analyzed in this work were taken while {\it PSP} was completely or predominantly outside the Alfv\'en critical zone \citep[depending on how this is defined; see Figure 9(a) of][]{RuffoloEA20,WexlerEA21}, so spatial domains with an isotropic aspect ratio are consistent with this view.}

In conclusion, we have defined and characterized the basic properties of contiguous domains of approximate magnetic pressure balance, i.e., nearly constant $|{\bf B}|$, which can be considered as Alfv\'enic domains. 
These domains relate to Alfv\'enic turbulence 
\citep{Barnes79a,Barnes81}
and dynamical plasma relaxation processes
\citep{ServidioEA08-depress}.
The domain duration relates to the waiting time between events that disrupt the near constancy of $|{\bf B}|$, many of which are switchbacks.  
The filling fraction of domains, which can serve as a measure of Alfv\'enicity of extended time periods, has a strong inverse relation with distance $r$ from the Sun over the first five {\it PSP} orbits.
The mean domain duration has a strong and continuous inverse relation with plasma $\beta$ over $0.1<\beta<2$.
Domain distributions have a broken power-law form with a break on the order of the turbulence correlation time, and are qualitatively consistent with previously measured waiting time distributions between switchbacks and magnetic PVI events. 
If domains represent statistically homogeneous spatial structures carried with the solar wind, their shapes are not highly elongated along the radial or Parker spiral direction.
These 
exercises in quantifying statistical 
properties of near-constant magnitude 
magnetic domains 
may help clarify what has been 
described as a flux tube or spaghetti model 
\citep{MccrackenNess66,Burlaga69,Borovsky08,Borovsky16}. 
A goal for the future is
to employ these 
clues and 
constraints to determine
the dynamical origin 
of this organized, cellular structure
of the solar wind magnetic field.

\acknowledgments 
This research has been supported 
 in part by grant RTA6280002 from Thailand Science Research and Innovation and the Parker Solar Probe mission under the 
 ISOIS project 
 (contract NNN06AA01C) and a subcontract 
 to University of Delaware from
 Princeton University (SUB0000165).
 Additional support is acknowledged from the  NASA LWS program  (NNX17AB79G) and the HSR program (80NSSC18K1210 \& 80NSSC18K1648).

%


%

\bibliography{domains}

\end{document}